\documentclass[journal=jacsat, manuscript=article, layout= twocolumn]{achemso}

\usepackage{etoolbox}

\usepackage{amsmath}
\usepackage{graphicx}
\usepackage{float}
\usepackage{physics}

\usepackage{lipsum}
\usepackage{xcolor}
\usepackage{acronym}

\makeatletter
\let\l@addto@macro\relax
\makeatother
\usepackage[fontsize=10pt]{scrextend}
\usepackage{setspace}

\newlength{\bibitemsep}
\newlength{\bibparskip}
\let\oldthebibliography\thebibliography
\renewcommand\thebibliography[1]{%
  \oldthebibliography{#1}%
  \setlength{\parskip}{\bibitemsep}%
  \setlength{\itemsep}{\bibparskip}%
}

\newacro{DBT}{Dibenzoterrylene}
\newacro{Ac}{Anthracene}
\newacro{NC}{nanocrystal}

\newcommand*{\sometext}{We perform laser spectroscopy at liquid helium temperatures (T=2\,K) to investigate single dibenzoterrylene (DBT) molecules doped in anthracene crystals of nanoscopic height fabricated by electrohydrodynamic dripping. Using high-resolution fluorescence excitation spectroscopy, we show that zero-phonon lines of single molecules in printed nanocrystals are nearly as narrow as the Fourier-limited transitions observed for the same guest-host system in the bulk. Moreover, the spectral instabilities are comparable to or less than one linewidth. By recording super-resolution images of DBT molecules and varying the polarization of the excitation beam, we determine the dimensions of the printed crystals and the orientation of the crystals' axes. Electrohydrodynamic printing of organic nano and microcrystals paves the way for a series of applications, where controlled positioning of quantum emitters with narrow optical transitions is desirable.} 

\let\oldmaketitle\maketitle
\let\maketitle\relax

\newcommand{\figref}[1]{Figure\,\ref{#1}}

\makeatletter
\renewcommand*{\acs@author@fnsymbol@symbol}[1]{
    \ifcase #1 *\or
    1\or
    2\or
    3\or
    4\or
    5\or
    6\or
    7\or
    8\or
    9\or
    10
    \fi
}
        
\renewcommand*\acs@contact@details{
    {\sffamily *\,E-mail: \acs@email@list }%
    \acs@number@list
}           

\patchcmd{\acs@address@list@auxii}
{\acs@author@fnsymbol{\acs@affil@marker@cnt}}
{\textsuperscript{\acs@author@fnsymbol{\acs@affil@marker@cnt}}}
{}{}

\patchcmd{\acs@address@list@auxii}
{{\acs@author@fnsymbol{\acs@affil@marker@cnt}\@nameuse{@altaffil@\@roman\@tempcnta}\par}}
{{\textsuperscript{\acs@author@fnsymbol{\acs@affil@marker@cnt}}\@nameuse{@altaffil@\@roman\@tempcnta}\par}}
{}{}        

\makeatother

\title{High-resolution cryogenic spectroscopy of single molecules in nanoprinted crystals}
\author{Mohammad Musavinezhad}
\affiliation{Max Planck Institute for the Science of Light, D-91058 Erlangen, Germany}
\alsoaffiliation{Department of Physics, Friedrich Alexander University Erlangen-Nuremberg, D-91058 Erlangen, Germany}
\author{Jan Renger}
\affiliation{Max Planck Institute for the Science of Light, D-91058 Erlangen, Germany}
\author{Johannes Zirkelbach}
\affiliation{Faculty of Physics, Ludwig-Maximilians–Universit\"at M\"unchen, D-85748 Garching, Germany.}
\author{Tobias Utikal}
\affiliation{Max Planck Institute for the Science of Light, D-91058 Erlangen, Germany}
\author{Claudio U. Hail}
\affiliation{California Institute of Technology, Pasadena, California 91125, USA}
\author{Thomas Basch\'e}
\affiliation{Department of Chemistry, Johannes Gutenberg-University, Mainz 55099, Germany}
\author{Dimos Poulikakos}
\affiliation{Institute for Mechanical Systems, ETH, Zurich, Switzerland}
\author{Stephan G\"otzinger}
\affiliation{Max Planck Institute for the Science of Light, D-91058 Erlangen, Germany}
\alsoaffiliation{Department of Physics, Friedrich Alexander University Erlangen-Nuremberg, D-91058 Erlangen, Germany}
\alsoaffiliation{Graduate School in Advanced Optical Technologies (SAOT), Friedrich Alexander
University Erlangen-Nuremberg, D-91052 Erlangen, Germany}
\author{Vahid Sandoghdar}
\affiliation{Max Planck Institute for the Science of Light, D-91058 Erlangen, Germany}
\alsoaffiliation{Department of Physics, Friedrich Alexander University Erlangen-Nuremberg, D-91058 Erlangen, Germany}
\email{vahid.sandoghdar@mpl.mpg.de}

\begin{document}

\twocolumn[
\begin{@twocolumnfalse}
\oldmaketitle
\begin{abstract}
\sometext
\end{abstract}
\end{@twocolumnfalse}
]


Molecules embedded in organic crystals hold great promise to act as solid-state emitters for applications in integrated quantum photonic circuits \cite{turschmann2019, toninelli2021}. In particular, individual dye molecules can possess lifetime-limited narrow electronic transitions in the order of 10\,MHz at liquid helium temperatures,\cite{moerner1989,orrit1990,toninelli2021} comparable to that of unperturbed alkali atoms. The fact that molecules can be synthesized with atomic precision and have a very small footprint of about 1\,nm, makes them ideally suited for a number of interesting applications. Indeed, single dye molecules are selectively addressable at cryogenic temperatures and have been utilized for the realization of a bright single-photon source,\cite{chu2017} achieving strong coupling and single-photon nonlinearities in a microcavity.\cite{pscherer2021} Some of the exciting future applications such as quantum networks require the collective coherent coupling of several quantum nodes to a common photonic mode.\cite{haakh2016,awschalom2021} However, efficient coupling of a large number of emitters will require a high degree of control over both their resonance frequencies and positions.

In an earlier work, we showed that \textit{p}-terphenyl nanocrystals (NCs) doped with a very small number of terrylene molecules could be printed by electrohydrodynamic dripping (EHD) and positioned with respect to photonic structures with subwavelength accuracy.\cite{hail2019} Moreover, we showed that terrylene molecules in this system were well protected and remained photostable even at room temperature. In this article, we extend our material system to that of dibenzoterrylene (DBT) in anthracene (Ac) and demonstrate that the zero-phonon lines (00ZPLs) connecting the ground state $\ket{g, \nu =0}$ and the lowest vibrational level of the excited state $\ket{e,\nu =0}$ in DBT molecules become nearly as narrow as the expected Fourier limit in bulk samples. This feature combined with subwavelength accuracy in positioning paves the way for large-scale coupling of molecules to photonic devices.

\section{Results and Discussion}
Electrohydrodynamic nanodripping is a versatile printing technique that utilizes an applied electric field at the tip of a micropipette to deposit individual sub-femtoliter droplets of an "ink" onto a surface. Upon evaporation of the solvent, particles or molecules coagulate to form nanostructures. This method allows for precise sub-wavelength placement of the nano-object and is applicable to a wide range of materials, including gold nanoparticles,\cite{galliker2012, schneider2015} quantum dots,\cite{kress2015} and organic molecules.\cite{hail2019} In this work, we explore the spectral behavior of DBT molecules in printed Ac crystals at cryogenic temperatures. DBT is a member of the polycyclic aromatic hydrocarbons (PAH) group and possesses exceptional spectral stability, strong 00ZPLs, high quantum efficiency,\cite{musavinezhad2023} and negligible dephasing when incorporated in suitable matrices at T$=2$\,K.\cite{nicolet2007linewidth, toninelli2021}

\begin{figure}[!t]
    \centering
    \includegraphics{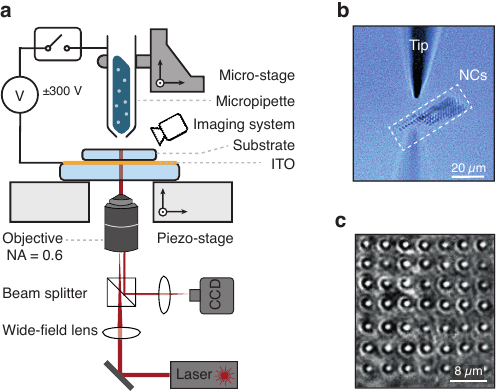}
    \caption{Electrohydrodynamic nanodripping for fabrication of organic crystals. (a) Sketch of the setup utilized for printing organic crystals. The setup integrates a low-resolution side channel and a high-resolution iSCAT channel for real-time imaging of printed structures during fabrication. (b) An example of the side-view for facilitating the coarse alignment of the micropipette (tip) above the sample substrate. The dashed rectangle marks the region of the printed array of nanocrystals (NCs). (c) An example of the iSCAT image of a printed array of Ac NCs.}
    \label{fig:setup}
\end{figure}

Principles of nanoprinting have been previously discussed, \textit{e.g.}, in Ref.\,[\hspace*{-4px}\citenum{galliker2012}]. \figref{fig:setup}a displays a schematic of the fabrication setup. In brief, the sample is placed on a transparent fused silica substrate with a 100\,nm indium-tin-oxide (ITO) coating that serves as a ground electrode. A gold-coated glass microcapillary nozzle (tip diameter: $1.5\pm0.5\,\mu$m) is filled with the printing ink. In this work, we used a 10:1 mixture of 1-octanol saturated with Ac, and $2\,\mu$M DBT in trichlorobenzene (TCB) as the ink (see Experimental Section). The tip is then placed approximately $5\,\mu$m above the sample surface, while monitored by a side-view imaging system (see \figref{fig:setup}b). A DC voltage is applied across the nozzle and the ground electrode to trigger the ejection of ink droplets. Periodic ejection of nano-droplets can be achieved at rates exceeding 100\,Hz, depending on the applied DC voltage and properties of the ink.\cite{galliker2012} Precise placement of the droplets is facilitated by piezoelectric positioners to adjust the sample relative to the nozzle. The nanocrystal growth is continuously monitored in real-time with an iSCAT microscope,\cite{taylor2019} as demonstrated in \figref{fig:setup}c.

As the solvent gradually evaporates, the ink droplet reaches supersaturation and the solute molecules precipitate to form crystals. Similar to other solution-based techniques, the quality of the printed crystals is significantly influenced by various factors including the choice of the solvent,\cite{chung2006} evaporation rate,\cite{hail2019} concentration, and the organic compound to be printed. Additionally, maintaining a limited dilution level of the ink is crucial to prevent clogging of the nozzle. Increasing the concentration of Ac above $\sim$7\,mg/mL may disrupt the ink flow and prevent the printing process. Moreover, a solvent with low surface tension is needed to enable ejection at voltages below the ionization threshold of the ambient atmosphere. In addition, a slower crystal growth is advantageous for achieving a high-quality molecular arrangement. The latter can be accomplished by using solvents with low evaporation rates at ambient conditions. It is also possible to increase the duration of crystal growth by maintaining the concentration of host molecules close to the saturation level. By doing so, the ink droplet reaches supersaturation as soon as evaporation starts and the growth time is merely limited by the full evaporation of the solvent. The solubility of 1-octanol for Ac reaches saturation at around 2.3\,mg/mL,
which falls well within the optimal concentration range for nanoprinting. Other favorable properties of this substance include low surface tension and slow evaporation
rate.

The size of the printed crystal can be adjusted by controlling the voltage amplitude and the dripping time. If, as for 1-octanol, the evaporation of the solvent is slow, consecutive nano-droplets can accumulate before the crystal is fully formed. Consequently, crystalline structures from hundreds of nanometers to several micrometers can be produced. In a first demonstration, we present a highly doped DBT:Ac microcrystal that is about 5\,$\mu$m wide and less than $500$\,nm in height. 

To conduct spectroscopic measurements on the embedded DBT molecules, the sample was cooled to 2\,K in a helium bath cryostat. The crystal was illuminated in the wide-field mode by a beam from a frequency-tunable narrow-band ($<$1\,MHz) Ti:sapphire laser. The back-reflected laser beam was blocked by a tunable long-pass filter, and the red-shifted fluorescence signal was measured with an avalanche photodiode detector (APD) or an EMCCD camera as the laser frequency was scanned. Device synchronization and measurement automation were done using a custom software based on pyLabLib.\cite{shkarin2023pll} A more detailed description of the optical setup is presented elsewhere.\cite{musavinezhad2023}

\begin{figure*}[t!]
    \centering
    \includegraphics{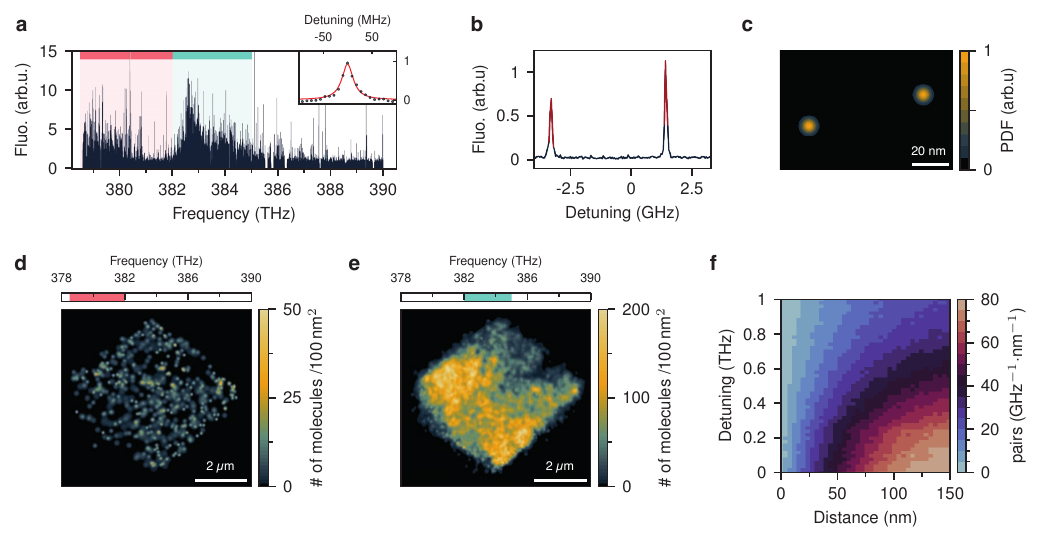}
    \caption{Spatial and spectral distribution of DBT  molecules in printed organic crystals. (a) Fluorescence excitation spectrum from $1\,\mu\text{m}^2$ of a printed DBT:Ac microcrystal recorded at 2\,K. The inset shows an example of a high-resolution spectrum of a single-molecule 00ZPL. The gaps in the scanning range are due to interruptions in the laser cavity lock. The two main insertion sites of DBT in Ac are marked with red and turquoise regions.(b) Spectra of two molecules located within a diffraction-limited spot. Red curves show the data points used to localize individual molecules. (c) Probability density function (PDF) of lateral positions for the two DBT molecules in (b). The location of each emitter is represented by a Gaussian spot. (d) Super-resolution image of 589 molecules between 378.5\,THz and 382\,THz, \textit{i.e}, the red region in (a). The embedded emitters mark the crystal's shape and boundaries. (e) Super-resolution image of 11197 molecules between 382\,THz and 385\,THz, \textit{i.e}, the turquoise region in (a). (f) Distribution of frequency detuning and distance between molecular pairs for emitters with less than 150\,nm separation.}
    \label{fig:microxtal}
\end{figure*} 

\figref{fig:microxtal}a presents the fluorescence excitation spectra of molecules within a region of 1\,$\mu$m$^2$ from a microcrystal recorded by the camera. Due to slight variations in the local environment, the transition frequencies of individual molecules vary within the inhomogeneous broadening (IB). We attribute the two spectral subpopulations to the two main insertion sites of DBT in Ac. A similar feature was previously observed in bulk Ac with sites at about 381.9 and 377.4\,THz.\cite{nicolet2007linewidth} For DBT:Ac, IB ranges from approximately 100\,GHz up to a few THz, depending on the crystalline quality and the insertion site.\cite{nicolet2007linewidth, pazzagli2018} 

If the doping level is not too high, the 00ZPLs of individual molecules do not overlap so that single molecules can be identified and addressed in frequency space.\cite{basche2008} The symbols in the inset of \figref{fig:microxtal}a show an example of a high-resolution scan through the 00ZPL of a single molecule, fitted with a Lorentzian function (red curve) of full width at half-maximum (FWHM), corresponding to $\gamma= (28 \pm 1)$\,MHz.

The large DBT doping level of the crystal allows us to establish a super-resolution image of the Ac crystal \cite{weisenburger2015}. As demonstrated in \figref{fig:microxtal}b, we can identify for two individual molecules through their non-overlapping 00ZPLs. The fluorescence point-spread function (PSF) of a molecule in a batch of frames is fitted by a two-dimensional (2D) Gaussian function, and its center is localized, determining the probability density function (PDF) of a molecule's lateral positions. The outcome yields a Gaussian spot, whereby its standard deviation ($\sigma$) is obtained from the localization precision. \figref{fig:microxtal}c illustrates the super-resolution image of the two molecules (same as in \figref{fig:microxtal}b) separated by $(69 \pm 4)$\,nm. 

\figref{fig:microxtal}d (2e) displays the super-resolution image of DBT molecules in the "red" ("blue") site, corresponding to transition frequencies below (above) 382 THz. We find that the molecules in the red site are nearly uniformly distributed. The great majority of the molecules (about 95\%), however, reside within the blue site and experience a more inhomogeneous distribution.

In \figref{fig:microxtal}f, we present a histogram of molecular pairs with less than 150\,nm separation to investigate the probability of coupling among two or more molecules, a phenomenon that can take place if molecular resonances and positions are very close to each other \cite{hettich2002, trebbia2022,lange2023}. We find that the crystal contains more than 1000 pairs of molecules that are separated by less than 10\,nm in the sample plane and at the same time, exhibit a frequency difference of less than 10\,GHz. However, we do not see any further correlation between the transition frequencies of the emitters and their spatial proximity. We note that the observed trend between frequency and distance in \figref{fig:microxtal}f is an artifact caused by the geometric growth factor $\propto 2\pi r$, where $r$ is the separation.

\begin{figure}[t!]
    \centering
    \includegraphics{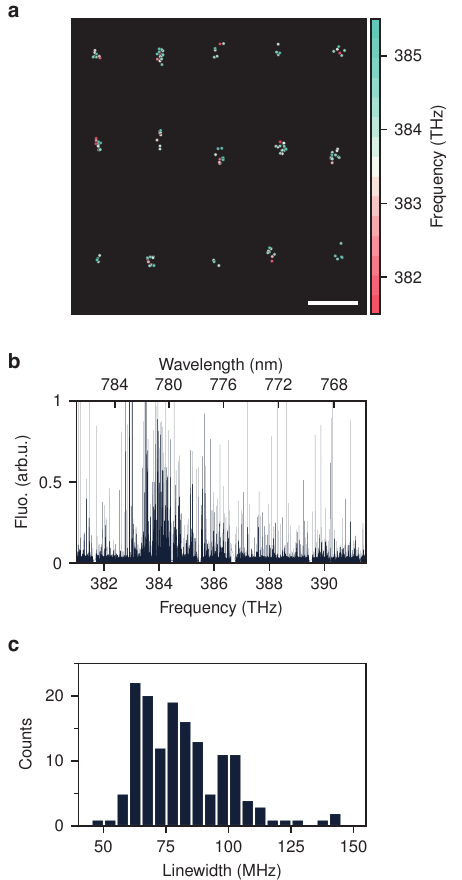} 
    \caption{Spectral properties of printed DBT:Ac nanocrystals. (a) Distribution of emitters in an array of printed nanocrystals. Each DBT molecule is shown as a point. Color code shows the resonance frequency. The scale bar shows $4\,\mu$m. (b) Fluorescence of DBT molecules in 100 NCs as a function of excitation frequency. (c) Linewidth distribution of 150 DBT molecules embedded in the printed Ac NCs.}
    \label{fig:printedNcs}
\end{figure}

\begin{figure*}[t]
    \centering
    \includegraphics{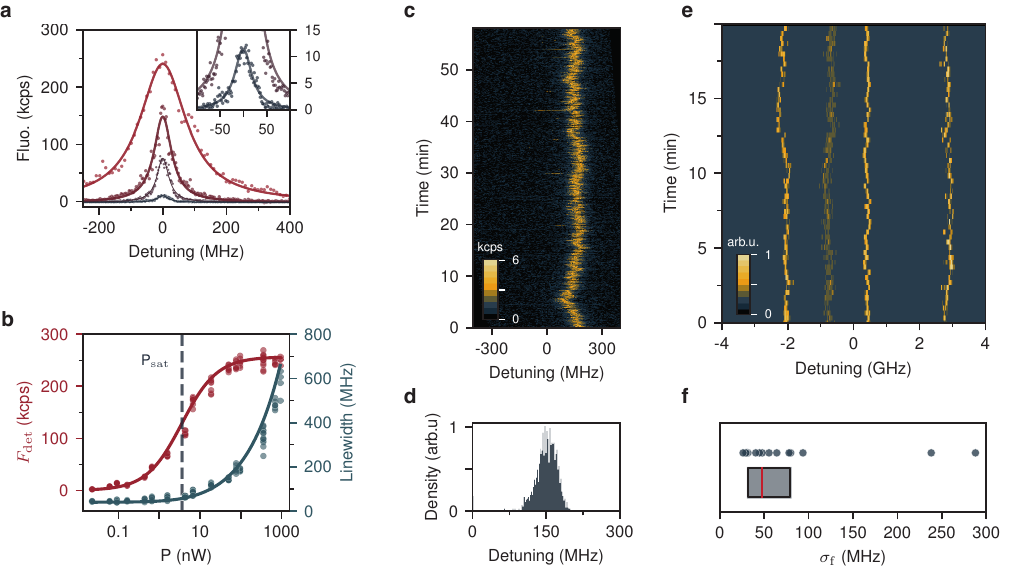}
    \caption{Saturation behavior and spectral stability of DBT molecules in printed nanocrystals. (a) Detected fluorescence as a function of laser detuning from the resonance of a single molecule, recorded at excitation powers $P= (0.17,\,1.6,\,6.6,\,95)$\,nW. Solid lines show Lorentzian fits. (b) Saturation of the fluorescence count rate and power broadening at high pump powers. Lines show theoretical fits with saturation power $P_\mathrm{sat}=3.6$ nW and maximum fluorescence rate of $F_{\infty}=257$ kcps and an intrinsic linewidth of $\gamma_0=41$ MHz. (c) Spectrum of a single molecule recorded at low excitation powers over one hour. (d) Histogram of the Lorentzian fit centers for 1600 scans in (c). The standard deviation of the measured values is $\sigma_f = 26$ MHz. (e) Spectra of 4 molecules recorded over 20 minutes. (f) Resonance frequency fluctuations $\sigma_f$ due to spectral diffusion for 15 DBT molecules in different NCs. The red line indicates the median at 48\,MHz. The box represents the range between 25 and 75 percentile of the population.}
    \label{fig:diffsat}
\end{figure*}

We now turn our attention to printed NCs. Due to their high surface-to-volume ratio, NCs typically sublimate within a few minutes following deposition under ambient conditions. To circumvent this, we applied a layer of poly(vinyl alcohol) (PVA) to the printed samples, facilitating their transfer to the cryostat under vacuum. \figref{fig:printedNcs}a presents an example of the 2D distribution of molecules in an array of DBT:Ac nanocrystals, whereby individual DBT molecules are illustrated as small circles color-coded according to their transition frequencies. 

\figref{fig:printedNcs}b displays the excitation spectra of 100 NCs. For a more quantitative analysis of the homogeneous linewidth, we performed laser frequency scans at a slower rate (700\,MHz/sec) over a frequency range of $383$ to $387$\,THz for at least $10$ repetitions. To minimize the effect of power broadening, the illumination intensity was set to $0.7~\mathrm{W}\cdot\mathrm{cm}^{-2}$, keeping the molecular excitation level well below saturation. Individual peaks in the excitation spectrum were then fitted with a Lorentzian function to extract their FWHM. \figref{fig:printedNcs}c presents a histogram of the measured linewidths ($\gamma$) for 150 molecules recorded in
at least 5 independent measurements. We measure molecular linewidths as low as about 47 MHz, which is slightly larger than the lifetime-limited linewidth of DBT in bulk Ac ($\sim 30$\,MHz).\cite{nicolet2007linewidth} 

A competitive feature of organic molecules as quantum emitters is their high emission rates and efficient interaction with light, which can also lead to a strong nonlinear response.\cite{maser2016,pscherer2021} Therefore, we investigated $\gamma$ and the detected fluorescence rate at resonance ($F_\mathrm{det}$) as a function of excitation power. \figref{fig:diffsat}a shows several excitation spectra of a molecule at different pump powers. As illustrated in \figref{fig:diffsat}b, both linewidth broadening and count rates align with the expected saturation law given by
\begin{align}
    F_\mathrm{det}(P)&=F_\mathrm{det}(\infty)\frac{P}{P+P_\mathrm{sat}} \quad \\ \quad \gamma(P)&=\gamma_0\sqrt{\frac{P+P_\mathrm{sat}}{P_\mathrm{sat}}},
    \label{Eq:sat}
\end{align}
where $\gamma_0$ is the intrinsic linewidth at weak excitation, and $P_\mathrm{sat}$ denotes the excitation power at saturation, \textit{i.e.}, $F_\mathrm{det}(P_\mathrm{sat}) =  F_\mathrm{det}(\infty) /2 $. The low value of $P_\mathrm{sat} = 3.6$ nW and the large number of detected photons $F_\mathrm{det}(\infty) = 257$ kcps (kilo counts per second) at full saturation are consistent with similar measurements conducted on DBT in bulk samples. This is also a strong indication that the quantum efficiency of the embedded molecules is similar to bulk.\cite{musavinezhad2023}

The remarkable sensitivity of molecular transitions to minute changes in their nanoscopic environment opens up possibilities for sensing a wide range of parameters such as strain,\cite{tian2014} electric fields and charges,\cite{shkarin2021} and temperature.\cite{esteso2023} However, accurate measurements of spectral changes require stable resonances. To investigate this in printed NCs, we repeatedly scanned the laser frequency over several individual 00ZPLs. \figref{fig:diffsat}c demonstrates the spectral stability of a single molecule  during the course of one hour, where the confocal excitation beam (0.6 nW) was kept fixed, and the red-shifted fluorescence was detected using an APD. The histogram of \figref{fig:diffsat}d shows the distribution of measured central frequencies with a standard deviation of $\sigma_f = 26$\,MHz, indicating that the resonance instability is comparable to one linewidth, which in this case amounted to $\gamma = 47$\,MHz.

We performed similar analyses for 15 molecules to gain more statistical information on the spectral stability of DBT in printed Ac NCs. An example of the excitation spectra for 4 molecules located in different NCs is presented in \figref{fig:diffsat}e. Only one of the emitters showed spectral jumps and all of them are photostable as long as the sample temperature remains below 30\,K. \figref{fig:diffsat}f summarizes the observed resonance instabilities $\sigma_f$ with a median value at approximately $48$\,MHz. Normalizing the spectral diffusion range of each molecule ($2\sigma_f$) by its linewidth yields a median value of $2\sigma_f/\gamma = 1.08$, underlining the molecules' good spectral stability.

\begin{figure*}[t]
    \centering
    \includegraphics{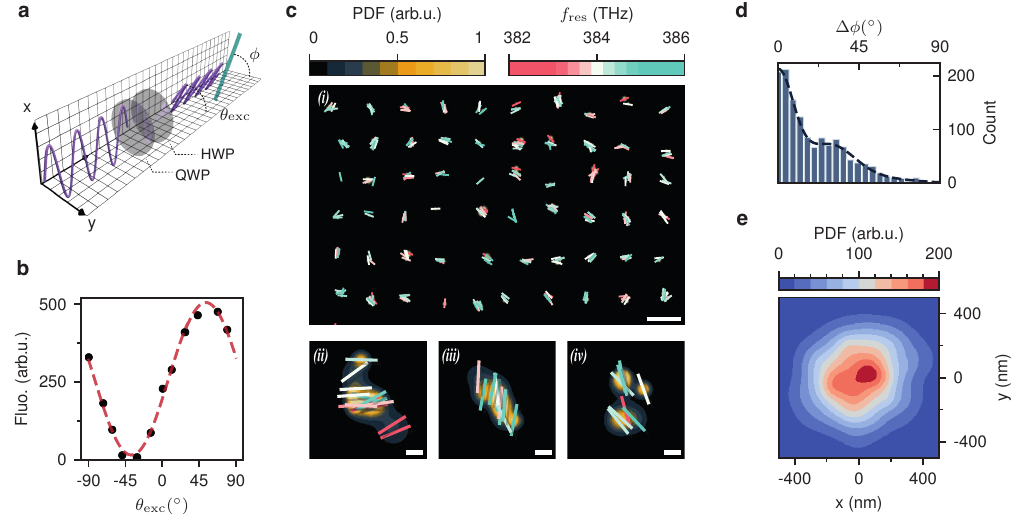}
    \caption{(a) Arrangement for determining the transition dipole moment orientation. A quarter-wave plate (QWP) and a half-wave plate (HWP) are used to adjust the excitation polarization $\theta_{\rm exc}$. $\phi$ indicates the angle of the transition dipole. (b) Fluorescence signal of a single molecule for different
$\theta_{\rm exc}$. The dashed line is a fit to a $\cos^2$ function.
    (c) \textit{i}: Simultaneous mapping of $\phi$ by frequency and polarization sweeps using wide-field illumination of an array of printed nanocrystals. Each molecule is represented by a short line, centered at its spatial coordinates and aligned parallel to its transition dipole angle $\phi$. The color of the lines encodes the corresponding resonance frequency. The scale bar shows $4\,\mu$m. 
    \textit{ii}-\textit{iv}: Close-ups for three exemplary NCs.
    The underlying image illustrates the PDF of the lateral position of the emitter, similar to \figref{fig:microxtal}e. The scale bars are 200\,nm.
    (d) Distribution of the angles between transition dipole moments $\phi$ in one NC. Dashed line shows a fit corresponding to two possible alignments of molecules, parallel or separated by $29^\circ$. (e) Spatial distribution of 261 DBT molecules with respect to the center of their corresponding NCs. The most likely orientation of DBT is set to the horizontal axis and can be associated with the \textit{b}-axis of Ac.}
    \label{fig:dipoles}
\end{figure*}

Low spectral diffusion and narrow resonances are good indicators of high crystalline quality of the host matrix. Moreover, the orientation of the DBT molecules can report on the crystallinity of the surrounding Ac matrix. In highly crystalline Ac samples, DBT molecules are predominantly incorporated along the \textit{b}-axis\cite{nicolet2007}. Thus, it is possible to deduce the orientation of the Ac crystal axis in the laboratory frame by determining the orientation of the transition dipole $\phi$ in the embedded DBT molecules. To achieve this, we rotated the polarization of the excitation beam $\theta_\mathrm{exc}$ and recorded the corresponding fluorescence count rate. \figref{fig:dipoles}a depicts the setup for this measurement. A half-wave plate (HWP) and a quarterwave plate (QWP) are used to adjust the polarization of the laser beam, whereby the angles are defined with respect to the optical table. By using the two wave plates, we were able to compensate for frequency dependence of
the final polarization introduced by the optical elements along the beam path. We note that the effect of Ac birefringence is negligible in our measurements because the thickness of the NCs is less than 100\,nm.

To minimize the uncertainty in the laser polarization angle $\Delta\theta_\mathrm{exc}$ caused by chromatic effects, we carefully calibrated the wave plates at three different frequencies, ensuring that $\Delta\theta_\mathrm{exc} < 1^\circ$. We conducted frequency scans in wide-field illumination for at least 10 different polarization angles. For each measurement, Lorentzian fits were used to determine the resonance frequency and fluorescence count rate of the spectra of single molecules. The count rates were then analyzed as a function of the polarization angle $\theta_\mathrm{exc}$ and fitted with a $\cos^2 \theta_\mathrm{exc}$ function because in the low-excitation regime, the molecular fluorescence intensity is proportional to the excitation intensity along its transition dipole, \textit{i.e.}, $F_\mathrm{det}\propto \cos^2(\phi - \theta_\mathrm{exc})$. \figref{fig:dipoles}b displays the results for one molecule. This fitting procedure allowed us to extract $\phi$ for each molecule. \figref{fig:dipoles}c(\textit{i}) presents an overview of the outcome for molecules from 60 NCs. Each molecule is represented by a short dash, centered at its location and aligned to its transition dipole orientation $\phi$. The color encodes the corresponding resonance frequency of the molecules (see scale on top). Close-ups of three NCs are displayed in \figref{fig:dipoles}c(\textit{ii}-\textit{iv}). The underlying image illustrates the PDF of the lateral position of the emitters, similar to the one shown in \figref{fig:microxtal}c. Here, the width of each Gaussian spot is determined from repeated measurements, reflecting the localization accuracy in long recordings.\cite{deschout2014} This accuracy is primarily constrained by small mechanical drifts in the course of several hours. 

The data in \figref{fig:dipoles}c present a clear trend in the orientation of the molecular transition dipoles. To quantify these correlations, we analyzed the data between pairs of DBT molecules located in the same NC. The histogram of differences ($\Delta\phi$) in the dipole moment orientation between molecules for about 7700 pairs is shown in \figref{fig:dipoles}d. We find that a significant fraction of the molecules are aligned with each other, and there are no pairs that are perpendicular to each other. However, we observe a secondary plateau in \figref{fig:dipoles}d. To model this behavior, we regarded $\phi$ as an independent random variable following a probability density $P_{\Phi}(\phi) =  a \mathcal{N}(\phi;0,\sigma_0) + (1-a)/2\,\mathcal{N}(\phi;\pm\phi', \sigma_1)$. Here, $\mathcal{N}(\phi;\theta_i,\sigma_i)$ represents a Gaussian distribution with mean $\theta_i$ and standard deviation $\sigma_i$. The first term indicates a population of $\Phi$ centered at 0, while the other terms define two symmetric sidebands at $\pm\phi'$ about this primary axis. Treating the remaining quantities as free fit parameters, we obtained the best estimation as $P_{\Phi}(\phi) = 0.63\,\mathcal{N}(\phi; 0^\circ, 6.0^\circ) + 0.37/2\, \mathcal{N}(\phi;\pm29^\circ,12.5^\circ)$. We attribute the population at $\phi'=\pm29^\circ$ to a second possible orientation for the insertion of DBT within the Ac NCs. The strong correlation observed in the alignment of DBT molecules provides compelling evidence for the single crystalline nature of the printed Ac, rather than a polycrystalline or amorphous structure where a large range of orientations would be possible. 

Next, we also analyzed the spatial distribution of 261 DBT molecules within 61 NCs. In each NC, we localized at least 3 molecules and determined their location and dipole orientation, as was shown in \figref{fig:dipoles}c. We assigned the center of each NC as the average values of the $x$ and $y$ coordinates for the emitters. Furthermore, we aligned the most common orientation of each NC along the $x$ axis. \figref{fig:dipoles}e displays the overall distribution obtained by adding the individual PDFs. We find a maximum molecule-molecule distance of approximately $850$\,nm. Considering that the finite PDF widths (approximately $20$\,nm) lead to a broadening of the distribution, we attribute $850$\,nm as the upper limit for the lateral size of the printed NCs. This result is in agreement with our AFM measurements on NCs that were printed with similar parameters. We verified that there was no correlation between orientation $\phi$ and location. 

Reduction of the crystal size can be envisioned via altering the substrate surface energy,\cite{markov2016} manipulating the wettability of both the substrate and printing nozzle,\cite{wu2016} and employing smaller tip diameters. Moreover, it is very important to counter the rapid sublimation of Ac NCs, \textit{e.g.}, by increasing Ac vapor pressure in a controlled environment. Lowering the substrate temperature in ambient conditions is less effective, particularly above the dew point.

\section{Conclusion}
Solid-state quantum emitters need a high degree of order in their environment because their intrinsically narrow optical transitions can be easily perturbed by atomic, molecular, or other nanoscopic dynamics in their surroundings, leading to frequency broadening and instability.\cite{schofield2022} This requirement is typically very difficult to satisfy when emitters are placed close to material interfaces, \textit{e.g.}, in thin films or in nanocrystals. In our current work, we have characterized micro and nanocrystals produced by electrohydrodynamic dripping with the advantage that individual crystals can be printed at a desired location, \textit{e.g.}, on a photonic circuit.\cite{hail2019} Since the crystal formation takes place within a short time of 1\,s during the printing process, one might be concerned with the resulting crystal quality. Our measurements show, however, that the printing process yields high-quality crystals. Here, we used the linewidth and frequency center of the spectra recorded from single DBT molecules embedded in anthracene nanocrystals as well as the orientation of their transition dipole moments as a measure. We found that we could reach linewidths that are less than twice broader than the Fourier limit of the bulk guest-host matrix and that spectral diffusion remains within the range of about one linewidth over hours.

The ability to print custom-designed thin micro and nanocrystals holds great promise for a number of applications such as coupling to open Fabry-Perot microcavities \cite{pscherer2021} or integrated photonic circuits \cite{rattenbacher2023,ren2022}. Moreover, the method can be used to form larger crystalline structures of different shapes, \textit{e.g.}, in the form of a waveguide or ring resonator. Future developments of the printing process as well as post-processing of the resulting crystals, \textit{e.g.}, through annealing, might improve the crystal quality further.

\section{Experimental Section} \label{sec:experimental}
\label{Methods}

\textbf{Nanoprinting.} To prepare the printing nozzles, borosilicate capillaries (World Precision Instruments TW100-4) are rinsed with acetone and then dried with nitrogen gas. The capillaries are pulled (Sutter Instrument P-2000) to form $\sim1.5\,\mu$m inner tip diameters. They are subsequently coated with 50\,nm titanium and 100\,nm gold by electron beam evaporation. Each sample is fabricated on a borosilicate glass coverslip (170$\,\mu$m thick) that was extensively cleaned by ultrasonic bath cleaning in diluted hexane and then deionized water for $30$ minutes each, followed by O$_2$ plasma cleaning.
To deposit one nanocrystal, a DC voltage of $300$\,V is applied to the nozzle for $1$ second. The substrate temperature is kept at $21^{\circ}$C during the fabrication. After the printing, a drop of $3$ wt\% poly(vinyl alcohol) in water is casted on the sample. We note that the faster sublimation of Ac NCs requires the polymer coating to be applied immediately after fabrication when the sample is still on the printing setup. 

\textbf{Printing Ink.} The ink is prepared by saturating 1-octanol with zone-refined Ac (30 passes; TCI chemicals) at around 2.3 mg/mL. DBT is dissolved in 1,2,4-trichlorobenzene (TCB) at 2\,$\mu$M concentration, and mixed with the Ac solution at a volume ratio of 1:10 TCB:octanol. In our previous work,\cite{hail2019} we used TCB as the solvent for terrylene:\textit{p}-terphenyl NCs which showed stable emission at room temperature. However, adapting the same recipe for DBT:Ac system did not show efficient integration of DBT in the Ac. The NCs printed using TCB as the solvent, exhibited a low density of DBT with unstable optical transitions.

\section{Acknowledgments}
This work was supported by the Max Planck Society, Deutsche Forschungsgemeinschaft (TRR 306 (QuCoLiMa)), and Bundesministerium für Bildung und Forschung (RouTe Project (13N14839)). We thank Alexey Shkarin for providing the lab control software and fruitful discussions.

\bibliography{Bibliography}

\end{document}